\documentclass[aps,pre,11pt]{revtex4}
\usepackage{amsthm}
\usepackage{newlfont}
\usepackage{graphicx}
\usepackage{subfigure}
\usepackage[section]{placeins}
\usepackage{psfrag}
\usepackage{color}
\usepackage{caption2}
\usepackage{flafter}
\usepackage{bm}%
\usepackage{amsthm}
\usepackage{newlfont}
\usepackage{graphicx}
\usepackage{subeqn}
\usepackage{subfigure}
\usepackage[section]{placeins}
\usepackage{psfrag}
\usepackage{caption2}
\usepackage{flafter}
\usepackage{bm}
\usepackage{bbding}
\usepackage{pifont}
\usepackage{wasysym}
\usepackage{amssymb}
\usepackage{dcolumn}
\usepackage{bm}
\begin{document}
\title{Impact of higher-order effects on dissipative soliton in metamaterials}
\author{A.K. Shafeeque Ali$^{1}$, Malik Zaka Ullah$^{2}$ and M. Lakshmanan$^{1}$}
 \address{1.Department of Nonlinear Dynamics, School of Physics, Bharathidasan University, Tiruchirappalli-620 024, India\\2.Department of Mathematics, Faculty of Science, King Abdulaziz University, Jeddah-21589, Saudi Arabia.}
\begin{abstract}
We study the influence of higher-order effects such as third order dispersion (TOD), fourth order dispersion (FOD), quintic nonlinearity (QN), self steepening (SS) and second order nonlinear dispersion (SOND) on the dynamics of dissipative soliton (DS) in metamaterials. Considering each higher-order effect as a perturbation to the system and following Lagrangian variational method, we demonstrate stable dynamics of DS as a result of the interplay between different higher-order effects. We also perform numerical analysis to confirm the analytical results.
\end{abstract}
\maketitle
\section{Introduction}
 Metamaterials allow engineering of material parameters from their basic constituents. This property helps one to alter various phenomena associated with the propagation of electromagnetic wave. In particular, dispersion in metamaterials can be tailored by stacking a number of highly dispersive sheets of metamaterials \cite{dis}. Then the desired nonlinear properties can be obtained either by using nonlinear insertions \cite{inser},
an element showing nonlinear response to resonant meta atoms or by embedding metamaterials into a nonlinear dielectric medium \cite{10}.
 Higher-order harmonics can be selectively generated by introducing varactor diodes into the split-ring resonator (SRR) circuit \cite{ilya}. The self steepening (SS) effect influenced by a light pulse depends on the size of the SRR element. The strength of birefringence can be tuned by adjusting the geometrical parameters of the constituent elements  \cite{imhof}. The band gap in the nonlinear metamaterial can be tuned by a variable capacitance insertion \cite{gorkunov}. A photometamaterial with meta-atoms containing both photodiodes and light emitting diodes can show breaking of inversion symmetry  at a particular incident wave intensity, which results in the emergence of second harmonic generation \cite{Maxim}. Tunable chirality in photonic metamaterials can be obtained with meta-molecule containing nonlinear nano- Au:polycrystalline indium-tin oxide layer sandwiched between two L-shaped nano-antennas \cite{yu}. Low threshold optical bistability at an ultralow excitation power can be observed in metamaterials with ultrathin holey metallic plates filled with nonlinear materials \cite{shi}.
 \par
 Soliton propagation is one of the striking and fascinating nonlinear phenomena \cite {ML} which has been investigated in the engineered metamaterials during the recent years. In metamaterials researchers have identified the existence of different kinds of solitons as a consequence of their engineering freedom. For example, existence of subwavelength discrete soliton due to the balance between tunneling of surface plasmon modes and nonlinear self-trapping has been identified \cite{yong}. Dispersive magnetic permeability provides controllability of the Raman soliton self-frequency shift \cite{yuan}. Bright-bright, dark-dark and dark-bright vector solitons can exist in the nonlinear isotropic chiral metamaterials \cite{Tsit}. Anomalous collision of elastic vector solitons has been identified in mechanical metamaterials as a result of large-amplitude characteristics of the solitons \cite{deng}. Nonlinearity induced wave transmission and generation of spatiotemporal solitons through an opaque slab with negative refraction have been studied \cite{nina}. Existence of low power gigahertz solitons have been identified in metamaterials through plasmon induced transparency \cite{Bai}. Metamaterials admit chirped bright, kink, and anti-kink quasi-solitons in the presence of SS effect \cite{anjan}.
 \par
On the other hand, a dissipative soliton is a stable localized structure formed by the double balance between nonlinearity and dispersion and between gain and loss which change the pulse energy. It can be observed in a variety of fields such as optics, cosmology, biology, condensed matter physics and medicine \cite{akh}. In fiber-laser cavities TOD can form stable and oscillatory bound states of DS \cite{malo1}. Stability of discrete dissipative localized modes in metamaterials composed of weakly coupled SRRs has been studied \cite{rosa}. Existence of knotted solitons, which are stable self localized dissipative structures in the form of closed knotted chains has been identified in magnetic metamaterials \cite{rosa2}. The delicate balance between input power and intrinsic losses results in the formation of stable dissipative breather in a superconducting quantum interface device metamaterials \cite{laza}.
 \par
 In contrast to the case of conventional positive index materials, the negative index metamaterials can show not only positive SS effect but also it can be negative, which is determined by the size of SRR contained in the meta-atom. As a result the solitons propagating inside the metamaterials either shift toward the leading edge or trailing edge depending upon the sign of the SS coefficient \cite{wen1}. Under the influence of negative SS effect the center of the pulse shifts toward the leading side and hence shifts a part of the energy of the pulse toward the blue side in contrast to the case of positive SS effect, where the center of the pulse shifts toward the trailing edge of the pulse (red side). Also the positive SS effect suppresses the self-frequency shift of  Raman soliton, whereas the negative SS effect enhances it \cite{yuan}. For many applications related to optic communication, stable dynamics of propagating light is necessary. Hence in this paper, we investigate the impact of higher-order effects such as third order dispersion (TOD), fourth order dispersion (FOD), quintic nonlinearity (QN) and second order nonlinear dispersion (SOND)  on the SS effect induced shift of the DSs. We consider each higher-order effect as a perturbation to the system and following Lagrangian variational method, we examine the possibility of stable propagation of the DSs in the metamaterials as a result of of the interplay between the above higher-order effects.
 \par
The paper is organized as follows. Following a self contained introduction, in Section II the theoretical model of the problem and variational analysis leading to the evolution equations of the pulse parameters are presented. In Section III, investigation on the impact of higher-order effects on the dissipative soliton is carried out in detail followed by a short presentation of the summary and conclusion in section IV.
 \section{Theoretical Analysis}
 The modified nonlinear Schr\"{o}dinger equation model which describes the propagation of electromagnetic waves in metamaterials is given by following normalized equation \cite{wen},
\begin{eqnarray}
\label{model}
\frac{\partial A}{\partial z } -i\sum_{m=2}^4 \frac{i^m \beta_m}{m!} \frac{\partial^m A}{\partial t^m }- i \gamma (|A|^{2}A) + i \gamma \xi (|A|^{4}A)+\gamma \sigma_1  \frac{\partial(|A|^{2}A)}{\partial t } \nonumber\\+i \gamma \sigma_2  \frac{\partial^2(|A|^{2}A)}{\partial t^2 } -g_l A =0,
\end{eqnarray}
where z and t are normalized propagation distance and time in a comoving frame of reference, respectively. In Eq. (\ref{model}) $A(z,t)$ stands for the normalized envelope of slowly varying electric field. $\beta_m$ and $\gamma$ are the $m^{th}$ order dispersion and nonlinearity coefficients, respectively. $\xi$ represents the quintic nonlinear coefficient. $\sigma_1$ and $\sigma_2$ are normalized first-and second-order nonlinear dispersion coefficients, respectively. Also $g_l=g-\alpha$, where g and $\alpha$ are the normalized gain and loss coefficients, respectively.
\par
Now we discuss the realization of properties presented in Eq. (\ref{model}) based on Drude-Lorentz model.
\begin{figure}[!ht]
\label{block}
\begin{center}
\subfigure[Array of metallic wires with radius of a wire 'r' and the distance between two wires 'a'.]{\label{wire}\includegraphics[height=4 cm, width=4 cm]{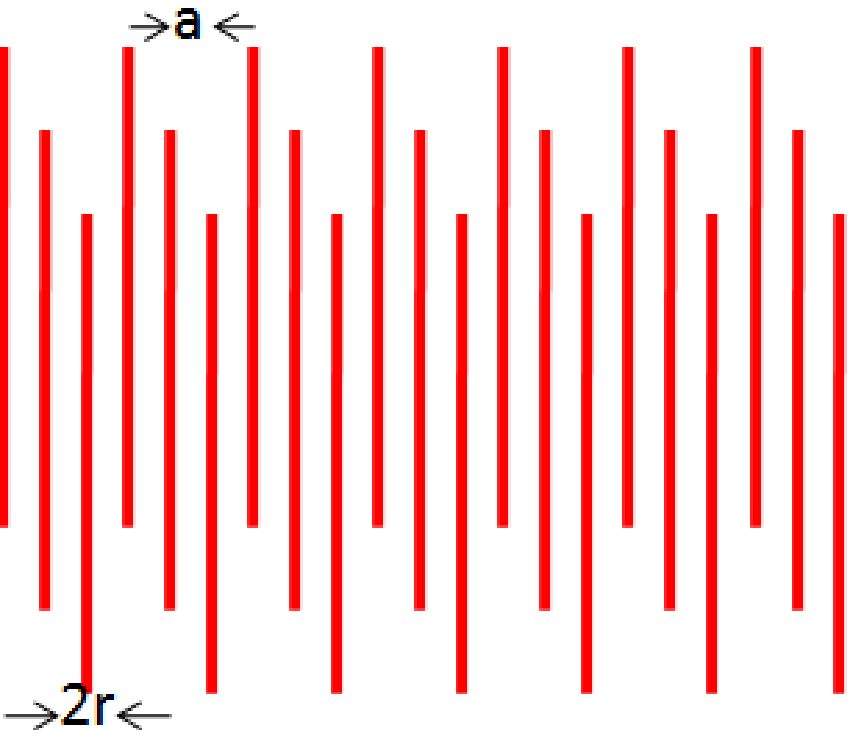}}~~~~
\subfigure[Split ring resonator with nonlinear insertions. The spacing between two rings and the radius  are  'd' and 'R' respectively.]{\label{ring}\includegraphics[height=4 cm, width=4 cm]{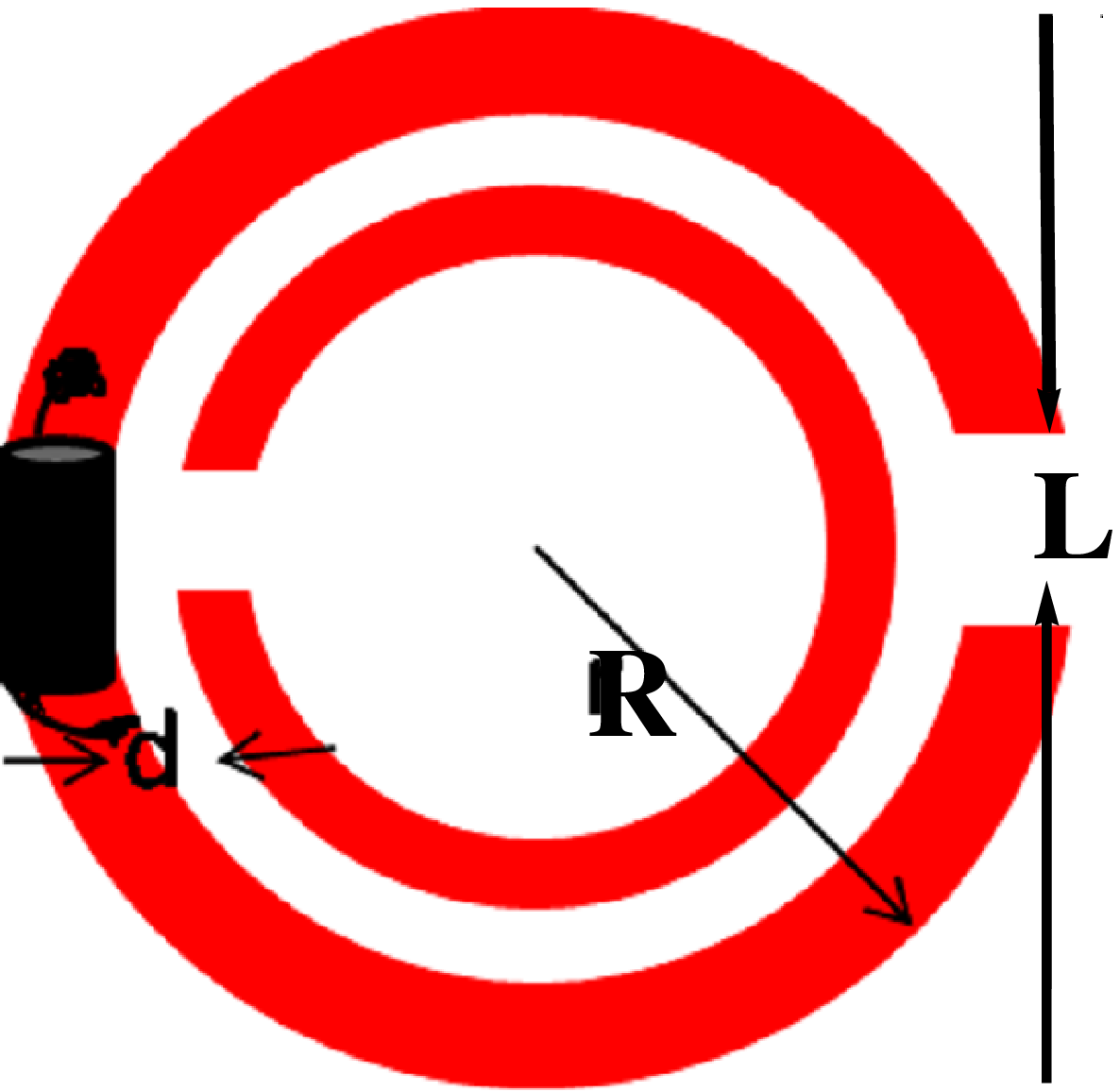}}
 \caption{Basic building blocks of a NIM}
 \end{center}
\end{figure}
According to Drude-Lorentz model \cite{Wooten}, the permittivity of a metal can be expressed as,
\begin{equation}
\epsilon(\omega)=1-\frac{\omega_{pe}^2-\omega_0^2}{\omega^2-\omega_0^2+i \omega \tau_c},
\end{equation}\\
where $\omega_0$ is the resonance frequency, $\omega_{pe}$ is the electron plasma frequency and $\tau_c$ is the damping frequency. For the particular case $\omega_0 < \omega < \omega_{pe}$, the electric permittivity is negative. In order to realize the structure exhibiting negative permittivity Pendry \emph{et al.} \cite{Pendry1} proposed a design of an array of metallic thin wires of radius $r$ and separation $a$ as shown in Fig. \ref{wire}, in which the value of plasma frequency  can be controlled by geometric parameters. The plasma frequency  of such an array of metallic thin wires is given by,
\begin{equation}
\omega_{pe}^2=\frac{2\pi c_0^2}{a^2 ln(a/r)},
\label{epl}
\end{equation}
where $c_0$ is the velocity of light. From Eq. (\ref{epl}) one can see that, the plasma frequency and hence the electric permittivity  depend on the geometrical parameters, namely the radius $r$ and spacing $a$ of the wires and thus on the structure of the lattice. In order to realize the negative magnetic response Pendry\emph{et al.} \cite{Pendry3} suggested that the split ring resonator structures, which consist of loops or ring structures, made of good conductors with a split or gap as shown in Fig. \ref{ring}. The effective magnetic permeability of the structure is given by,\\
 \begin{equation}
\mu(\omega)=1-\frac{F \omega^2}{\omega^2-\omega_0^2+i \tau \omega},
\label{permi}
\end{equation}
where $\omega_0=\frac{3 L c_0^2}{\pi R^3 ln(2 \omega/d)}$ is the resonant frequency, $F=\frac{\pi R^2}{L^2}$ is the volume filling factor and $\tau$ is the damping coefficient. Here again the resonant frequency, damping factor, filling factor and hence the effective permeability are functions of geometrical parameters of the structure. Thus, the thin wire lattice and split ring resonator are the two basic building blocks of NIM
medium where the former gives the negative electric responses and the latter gives the negative magnetic responses. Now various coefficients of Eq. (\ref{model}) can be represented as \cite{wen, R22, R21},
$\beta_2=\frac{1}{c_0 \omega_0 n}(1+\frac{3\omega_{pm}^2 \omega_{pe}^2}{\omega_0^4})-\frac{1}{c_0 \omega_0 n^3}(1-\frac{\omega_{pm}^2 \omega_{pe}^2}{\omega_0^4})^2$, $\beta_3=-\frac{12 (\omega_{pm}/\omega_{pe})^2}{n c \omega_{pe}^2 (\omega/\omega_{pe})^6}-\frac{3\beta_2((\omega /\omega_{pe})^4 - (\omega_{pm} /\omega_{pe})^2)}{3\omega_{pe}n^2(\omega/\omega_{pe})^5}$, $\beta_4=\frac{60 n^2 \omega_{pe}^2 \omega_{pm}^2}{k_0 \omega^6 -\frac{3\beta_2^2}{k_0}}$, $\xi=\frac{1}{2n \omega^2 T_p^2} |\frac{1}{n}(1+\frac{3\omega_{pe}^2 \omega_{pm}^2}{\omega_0^4})-\frac{(1-\frac{\omega_{pe}^2 \omega_{pm}^2}{\omega_0^4})^2}{n^3}|$, $\sigma_1=\frac{1}{\omega T_p}(1+\frac{\omega_{pm}^2 \omega_{pe}^2-\omega_0^4}{n^2 \omega_0^4}-\frac{\omega_{pm}^2+\omega_0^2}{\omega_{pm}^2-\omega_0^2})$ and $\sigma_2=\frac{1}{\omega^2 T_p^2}(\frac{\omega_0^2}{\omega_0^2-\omega_{pm}^2}-\frac{1}{4n^2}(1+\frac{3\omega_{pm}^2\omega_{pe}^2}{\omega_0^4}+\frac{1}{4n^4} (1-\frac{\omega_{pe}^2\omega_{pm}^2}{\omega_0^4})^2))$, where $n=\sqrt{\epsilon(\omega) \mu(\omega)}$ is the index of refraction. $\omega_{pm}$, $k_0$ and $T_p$ are the magnetic plasma frequency, propagation constant and initial pulse width, respectively. Note that the choice of the above parameters depends upon the selection of $\omega_{pm}$ and $\omega_{pe}$ that will determine the engineered size of the constituents of the NIM structures.
\par
Now, let us rewrite Eq. (\ref{model}) in the form of a perturbed nonlinear Schr\"{o}dinger equation,
\begin{eqnarray}
\label{per}
i \frac{\partial A}{\partial z } -\frac{\beta_2}{2}\frac{\partial^2 A}{\partial t^2}+ \gamma (|A|^{2}A)= i \vartheta(A),
\end{eqnarray}
where $\vartheta(A)$ is defined as,
\begin{eqnarray}
\label{per1}
\vartheta(A)= \frac{\beta_3}{6}\frac{\partial^3 A}{\partial t^3}+ i \frac{\beta_4}{24}\frac{\partial^4 A}{\partial t^4}- i \gamma \xi (|A|^{4}A)-\gamma \sigma_1  \frac{\partial(|A|^{2}A)}{\partial t }\nonumber\\ -i \gamma \sigma_2  \frac{\partial^2(|A|^{2}A)}{\partial t^2 }+g_l A.
\end{eqnarray}

Now, we follow the Lagrangian variational method developed by Anderson to study the characteristics of soliton evolution \cite{an1,an2}. The Lagrangian density corresponding to Eq. (\ref{per}) can be written as follows,
\begin{eqnarray}
L= \frac{i}{2}\,(A\frac{\partial A^*}{\partial z }- A^* \frac{\partial A}{\partial z })-\, \frac{\beta_2}{2}  |\frac{\partial A}{\partial t }|^2 -\frac{\gamma}{2}\,|A|^{4}+i(\vartheta A^* -\vartheta^* A).
\end{eqnarray}
As the problem deals with DSs let us now choose the Pereira-Stenflo solution \cite{pss} as ansatz for the solitary pulse,
\begin{eqnarray}
\label{ans}
A(z,t)=A_0(z)[sech(\rho(z)(t-t_p(z)))]^{(1+i w(z))} e^{i (\varphi (z)- \delta(z)(t-t_p(z)))},
\end{eqnarray}
where the parameters $A_0$, $\rho$, $t_p$, $w$, $\varphi$ and $\delta$ are assumed to be functions of the propagation distance (z). $A_0$ and $\rho$ define the amplitude and half-width, respectively, of the soliton, whereas $t_p$, $w$, $\varphi$ and $\delta$ are temporal position, frequency
chirp, phase and frequency shift, respectively. The variations of these parameters during the evolution of soliton can be found using Lagrangian variational analysis. The reduced Lagrangian of the system can be calculated by using the following relation,
\begin{equation}
\label{leg1}
\langle L\rangle=\int_{-\infty}^\infty L dt.
\end{equation}
After the substitution of $A(z,t)$ and its various derivatives on the right side of Eq. (\ref{leg1}) inside the integral and performing necessary integrations one can obtain the reduced Lagrangian as follows,
\begin{eqnarray}
\label{eg}
\langle L\rangle=2\frac{A_0^2}{\rho}(\frac{\partial \varphi}{\partial z }+\delta \frac{\partial t_p}{\partial z})-w\frac{A_0^2}{\rho^2} \frac{\partial \rho}{\partial z}+N \frac{A_0^2}{\rho}\frac{\partial w}{\partial z}+\rho \frac{A_0^2}{3}(1+w^2)\\ \nonumber+\frac{A_0^2}{\rho}(\delta^2-\frac{2}{3}A_0^2)+i \int_{-\infty}^\infty (\vartheta A^* -\vartheta^* A)dt,
\end{eqnarray}
where $N=\ln(2)-1$, $\beta_2=-1$ and $\gamma=1$. Varying the above effective Lagrangian with respect to the variational parameters given in Eq. (\ref{ans}) and performing the integration after substituting $\vartheta$ and $\vartheta^*$ from Eq. (\ref{per1}), we get the following system of six coupled nonlinear evolution equations corresponding to the six variational parameters,
\begin{subequations}
\label{evolu}
\begin{eqnarray}
\label{evolu1}
\frac{d t_p}{d z }= 2 t_p(g_l+\frac{4}{15}\gamma \sigma_2 A_0^2 w \rho^2)- \delta,
\end{eqnarray}
\begin{eqnarray}
\label{evolu2}
\frac{d \rho}{d z }= \rho (\frac{2}{15} w \rho^2 (5 -A_0^2(8 \gamma \sigma_2  +4 N \gamma \sigma_2 ))-2N g_l ),
\end{eqnarray}
\begin{eqnarray}
\label{evolu3}
\frac{d A_0}{d z }= w \rho^2(\frac{1}{3} A_0 -\frac{4}{5}A_0^3\gamma \sigma_2(1 -\frac{N}{3}))-ln(2)A_0 g_l,
\end{eqnarray}
\begin{eqnarray}
\label{evolu4}
\frac{d \delta}{d z }= \frac{4}{15}(2 \gamma \sigma_1 A_0^2 w \rho^2-15 ln(2)\delta g_l+w \rho^2 \delta(5-\gamma \sigma_2 A_0^2 (16-4 N ))),
\end{eqnarray}
\begin{eqnarray}
\label{evolu5}
\frac{d\varphi}{d z }= \frac{1}{4 f N_1}(\frac{4f A_0^2}{3}(2-\acute{N}-\frac{8\gamma \xi A_0^2}{5})+4 f w g_l(2+N-\acute{N}-N \acute{N}-2ln(2)N_1) \nonumber \\-\frac{2}{3}A_0 \rho(1 +\acute{N}- w^2+3 \acute{N} w^2)+\frac{8}{5} A_0^3 \rho \gamma(\sigma_2+\sigma_2  w^2(\frac{2}{3} N -  1 +2 \acute{N}-\frac{2 N \acute{N}}{3})-\frac{4 \acute{N} \sigma_1 t_p w }{3})\nonumber\\
+\beta_4 A_0 \rho^3(\frac{7}{90} +\frac{1}{9} w^2 +\frac{1}{30}  w^4)+\frac{1}{3}A_0 \rho (1+w^2) (2\beta_3 \delta+ \beta_4 \delta^2)+f(\frac{2 \beta_3  \delta^3}{3}+\frac{\beta_4  \delta^4}{6})\nonumber\\
-8  t_p f \delta g_l N_1 -\frac{32}{15} \gamma \sigma_2 t_p A_0^3 w \rho \delta N_1 +2 f \delta^2 N_1
+\frac{8}{3} \gamma f A_0^2 (\sigma_2+\sigma_1 \delta )),
\end{eqnarray}
\begin{eqnarray}
\label{evolu6}
\frac{d w}{d z }= 2 ln(2)g_l N_1 w-\frac{2}{3} N_1 A_0^2+16 t_p \delta g_l N_1+\frac{1}{4(N-1)N_1}( \frac{8A_0^2}{3}(\acute{N}-\frac{8}{3} A_0^2 \gamma\rho-2)\nonumber\\ +8g_l w ((\acute{N}-N)(1-2ln(2))-2+N\acute{N})  +\rho^2(\frac{4}{3}(1+\acute{N}-w^2(1-3\acute{N}))+\frac{16A_0^2\gamma}{5}(\acute{N} \sigma_1 t_p w \nonumber\\ +\sigma_2 w^2 (1-\frac{2}{3}N^2-\frac{2}{3}N+\frac{2}{3}N \acute{N} )-1)) -\frac{\beta_4 \rho^4}{3}(\frac{7}{15}+\frac{2w^2}{3}+\frac{w^4}{5})\nonumber\\ -\frac{16}{3}\sigma_1 A_0^2\delta-\frac{2}{3}\rho^2 \delta(1+w^2 \frac{\delta^2}{\rho^2})(\delta \beta_4+2 \beta_3) +\frac{64}{15} \sigma_2 \gamma t_p A_0^2 w \rho^2 \delta (1+N+ \frac{4\delta}{5t_p \rho^2 w}) +4\delta^2 N_1)\nonumber\\-g_l \frac{4 t_p \delta}{N-1}+\frac{\rho^2}{N-1}(3+w^2+8\sigma_2\gamma A_0^2 w^2 \rho^2(N-\frac{1}{5}+ \frac{2t_p \delta}{15 w}+ \frac{2\sigma_1 t_p}{15\sigma_2 w})+ \frac{\delta^2}{\rho^2}),
\end{eqnarray}
\end{subequations}
where $f=\frac{A_0}{\rho}$, $\acute{N}=\frac{N}{N-1}$ and $N_1=1-\acute{N}$. The above set of six coupled nonlinear differential equations given in Eq. (\ref{evolu}) provide the information about the variation of parameters given in Eq. (\ref{ans}) during the evolution of the dissipative soliton under the influence of perturbations defined in Eq. (\ref{per1}). Now we solve the set of equations given in Eq. (\ref{evolu}) numerically to know the behavior of soliton parameters during the evolution. We consider the input soliton is an unchirped and unit amplitude pulse with  zero initial phase, temporal position and frequency shift. Also we assume the metamaterial in which the soliton propagates is a nonlinear, dissipative and effective gain medium for compensating loss \cite{gan1, gan2}. Hence we choose $A_0(0)=1$, $\rho(0)=1$, $t_p(0)=0$, $w(0)=0$, $\varphi(0)=0$, $\delta(0)=0$ $g=0.15$, $\alpha=0.1$ and $\gamma=1$ to solve the system of equations.
 \section{Result and Discussion}
 Here we present the outcome of the variational analysis based on the coupled nonlinear differential equations given in Eq. (\ref{evolu}) and the numerical results after solving Eq. (\ref{model}). We adopt standard split-step Fourier method for our numerical simulation. The parameter values used are $g=0.15$, $\alpha=0.1$, $\gamma=1$, $\beta_2=-1$ and  $A_0(0)=1$.  Initially we study the influence of SS effect on the DS propagation. Then we examine the impact of other linear and nonlinear higher-order effects on the SS induced temporal shift. Finally, we demonstrate stable dynamics of the DS as a result of the interplay between various higher-order effects.
 \subsection{Impact of self steepening}
 \begin{figure*}[htb]
    \subfigure[]{\label{1a}\includegraphics[height=5 cm, width=7 cm]{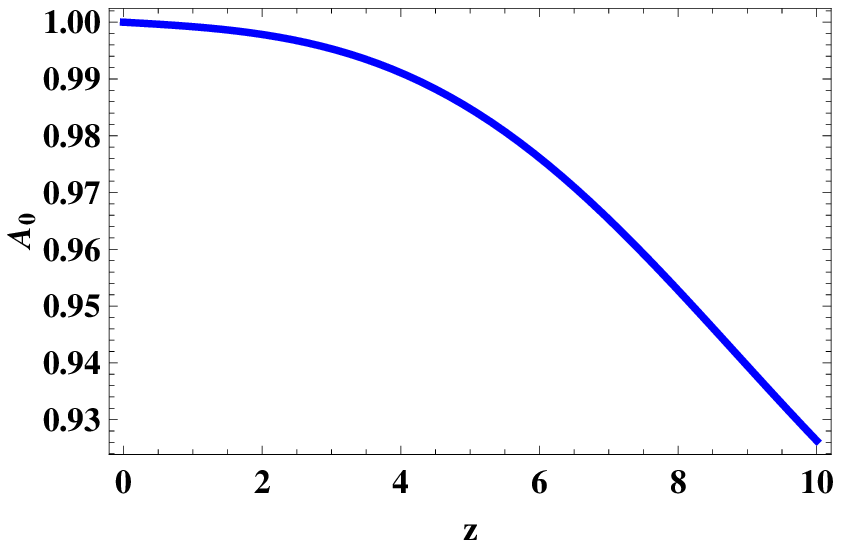}}
    \subfigure[]{\label{1b}\includegraphics[height=5 cm, width=7 cm]{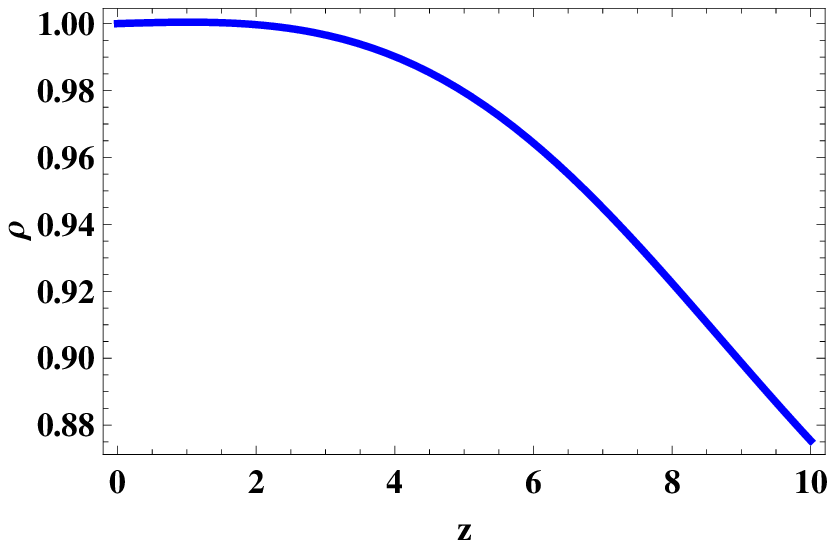}}
    \subfigure[]{\label{1c}\includegraphics[height=5 cm, width=7 cm]{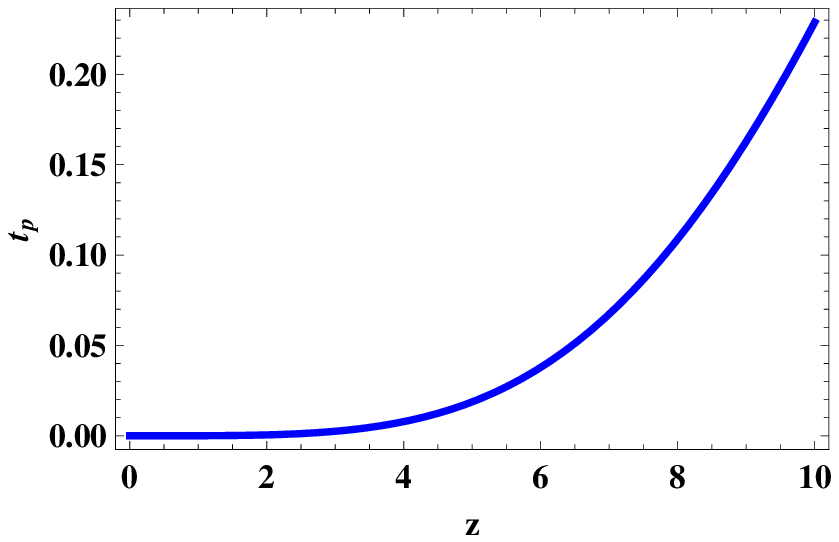}}
    \subfigure[]{\label{1d}\includegraphics[height=5 cm, width=7 cm]{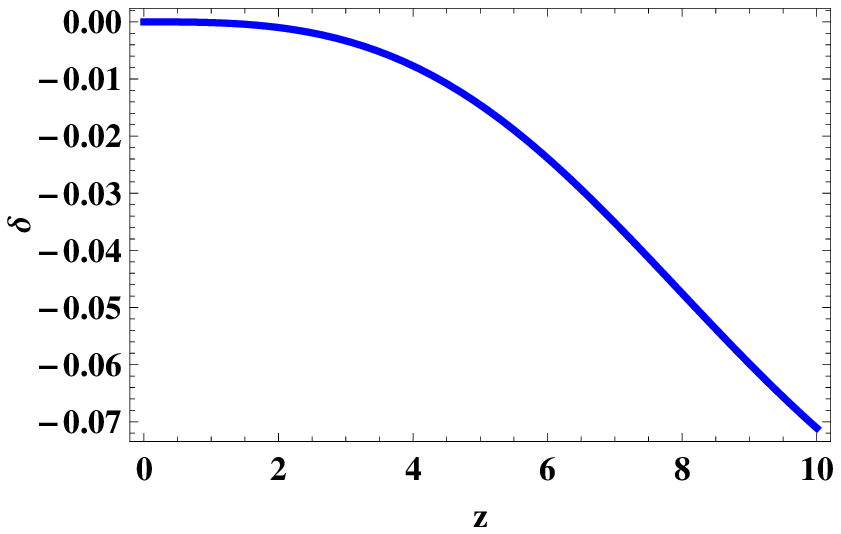}}
    \subfigure[]{\label{1e}\includegraphics[height=5 cm, width=7 cm]{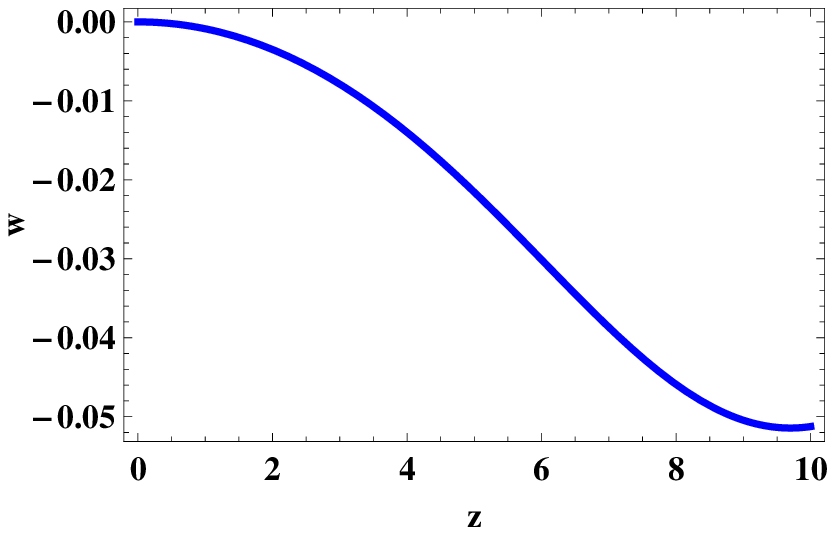}}
    \subfigure[]{\label{1f}\includegraphics[height=5 cm, width=7 cm]{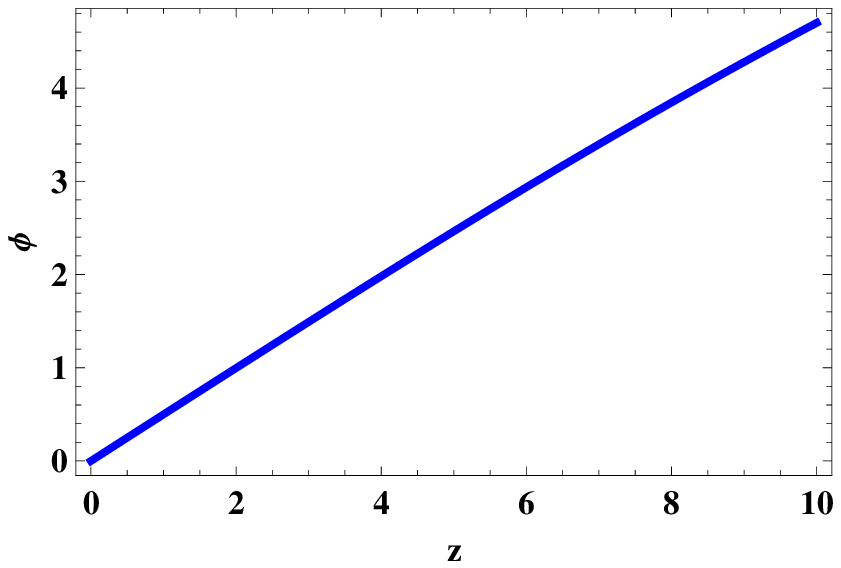}}
      \caption{(Color online.) Changes in different variational parameters under the influence of positive SS effect ($\sigma_1=0.5$) during the propagation of soliton.}
  \label{fig1}
\end{figure*}
Now we discuss the influence of the SS effect on the propagation of DS in the metamaterials in detail. In contrast to the case of conventional positive index materials, the negative index metamaterials can show not only positive SS effect but also it can be negative, which is determined by the size of SRR circuit contained in the meta-atom. We initially discuss the results from the variational analysis results based on Eq. (\ref{evolu}). Fig. \ref{fig1} depicts the changes in different variational parameters under the influence of positive SS effect ($\sigma_1=0.5$) during the propagation of soliton. The pulse energy, $E=\int_{-\infty}^\infty |A|^{2} dt$, should be a conserved quantity during the evolution of the soliton. From Eq. (\ref{eg}), the cyclic nature of the variational parameter $\varphi(z)$ gives,
\begin{eqnarray}
\label{sd}
E=2\frac{A_0^2}{\rho}.
\end{eqnarray}
 Figs. \ref{1a} and \ref{1b} show the changes in the parameters $A_0$  and $\rho$. It is clear from the figures that the parameters vary in such a way to keep the  total pulse energy given by Eq. (\ref{sd}) a constant during the evolution. Variations of the temporal position $t_p$ and the frequency shift $\delta$ are depicted in Figs. \ref{1c} and  \ref{1d}, respectively. Fig. \ref{1c} shows that as the soliton evolves the value of the temporal position increases. Hence the soliton peak shifts toward the trailing side of the pulse. Consequently $\delta$ decreases (Fig. \ref{1d}) and as a result the soliton spectrum shifts toward the low energy red side. Hence one can conclude that due to the influence of positive SS effect the peak of the DS shifts toward the trailing side and hence the spectrum shifts toward the red side when it propagates in the metamaterials. Also Figs. \ref{1e} and \ref{1f} depict the changes in the frequency chirp and the phase, respectively, during the evolution of the soliton.
 \begin{figure*}[htb]
    \subfigure[]{\label{2a}\includegraphics[height=5 cm, width=7 cm]{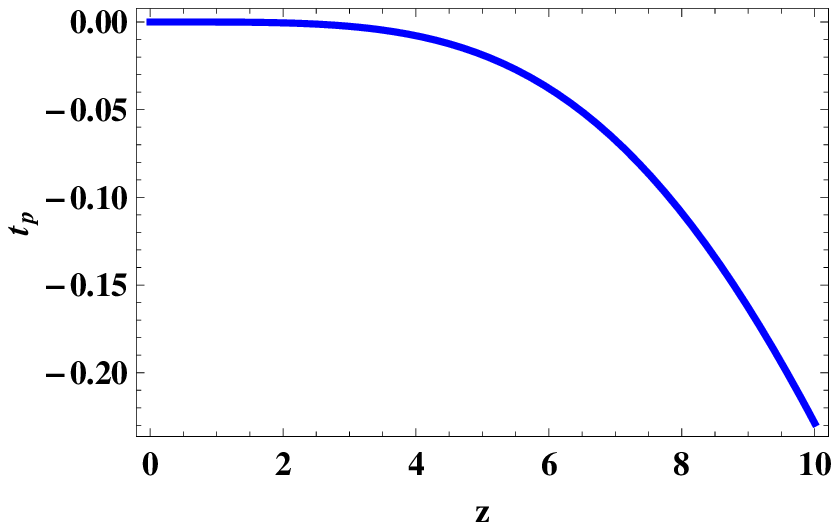}}
    \subfigure[]{\label{2b}\includegraphics[height=5 cm, width=7 cm]{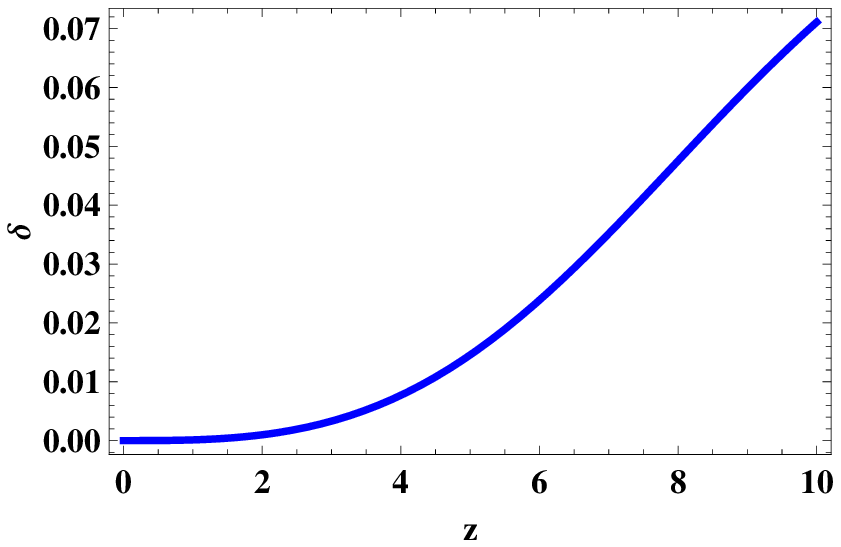}}
      \caption{(Color online.) Impact of negative SS effect ($\sigma_1=-0.5$) on temporal position and frequency shift during the propagation of soliton.}
  \label{threecore_normal}
\end{figure*}
On the other hand when the SS effect is negative the evolution of the DS is quite different as compared to the positive SS case.  Figs. \ref{2a} and \ref{2b} show the variations of the temporal position and frequency shift, respectively, when the SS effect is negative. As compered to the case of positive SS effect, here the changes are opposite in nature. Here the temporal position decreases and the frequency shift increases during the evolution. Hence the peak of the DS shifts toward the leading side of the pulse and hence the spectrum shifts toward the high energy blue side when the SS effect is negative.
 \begin{center}
 \begin{figure*}[htb]
 \subfigure[]{\label{3a}\includegraphics[height=6 cm, width=6 cm]{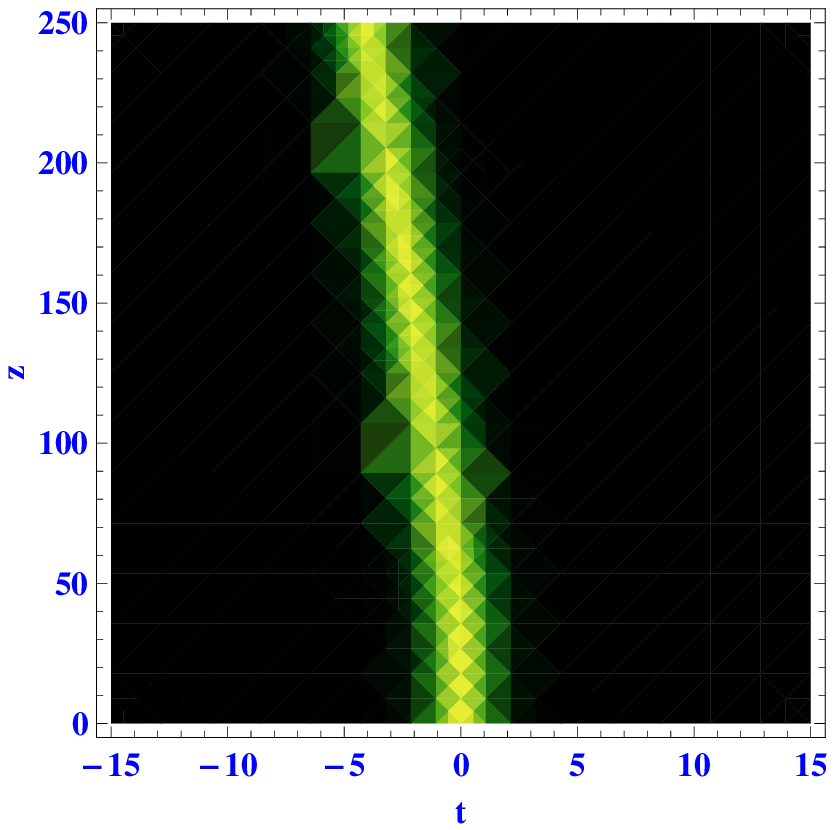}}
\subfigure[]{\label{3b}\includegraphics[height=6 cm, width=6 cm]{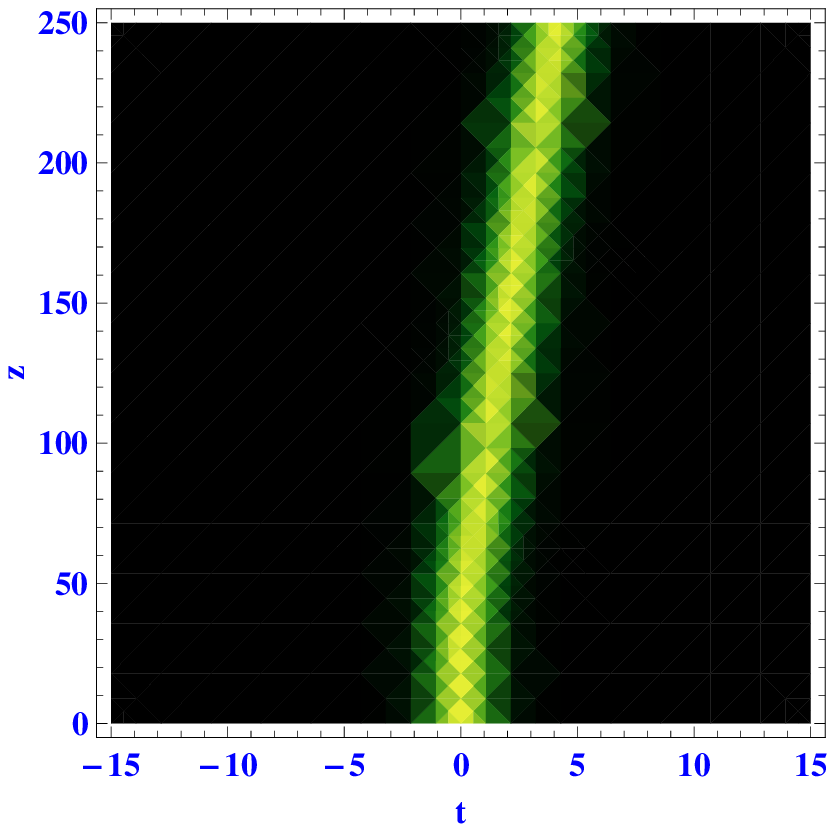}}
\subfigure[]{\label{3c}\includegraphics[height=6 cm, width=4.5 cm]{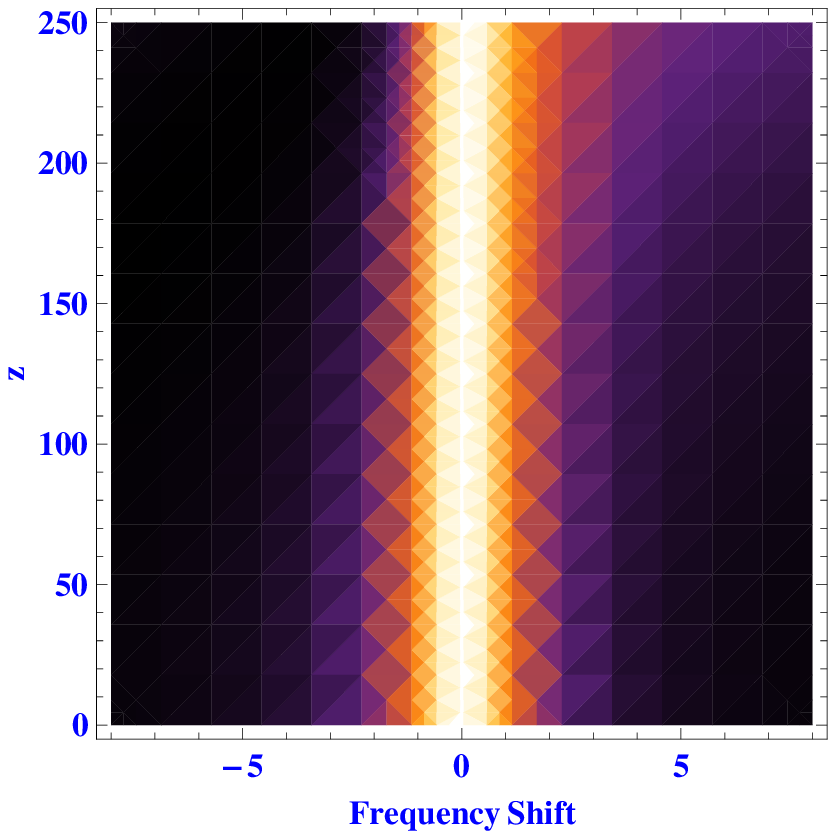}}~~~~~~
\subfigure[]{\label{3d}\includegraphics[height=6 cm, width=4.5 cm]{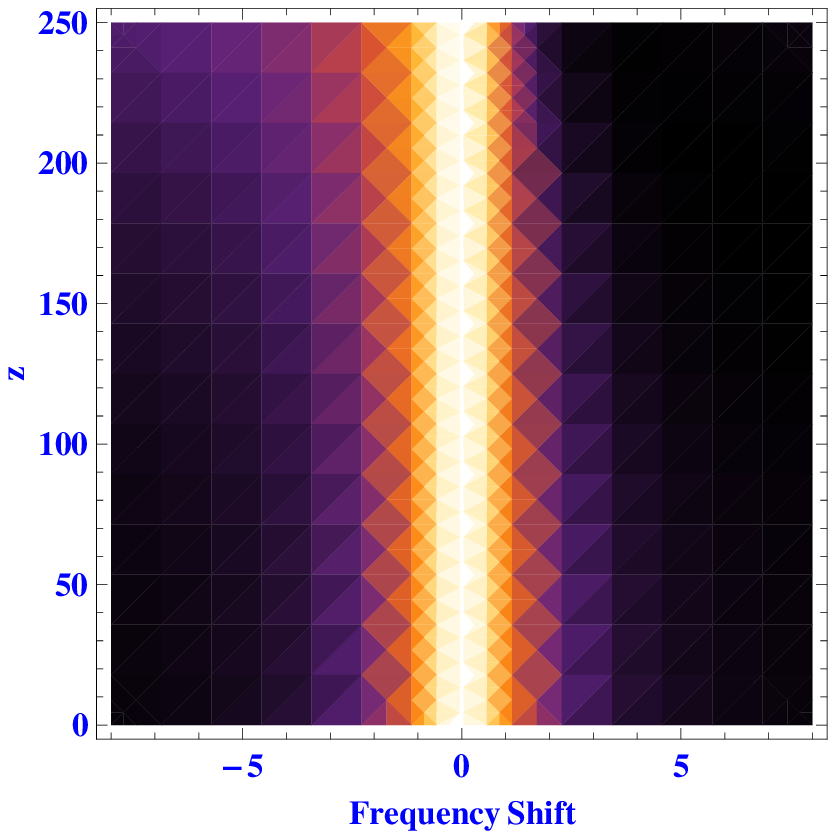}}~~~~~
\subfigure {\label{3e}\includegraphics[height=4 cm, width=1 cm]{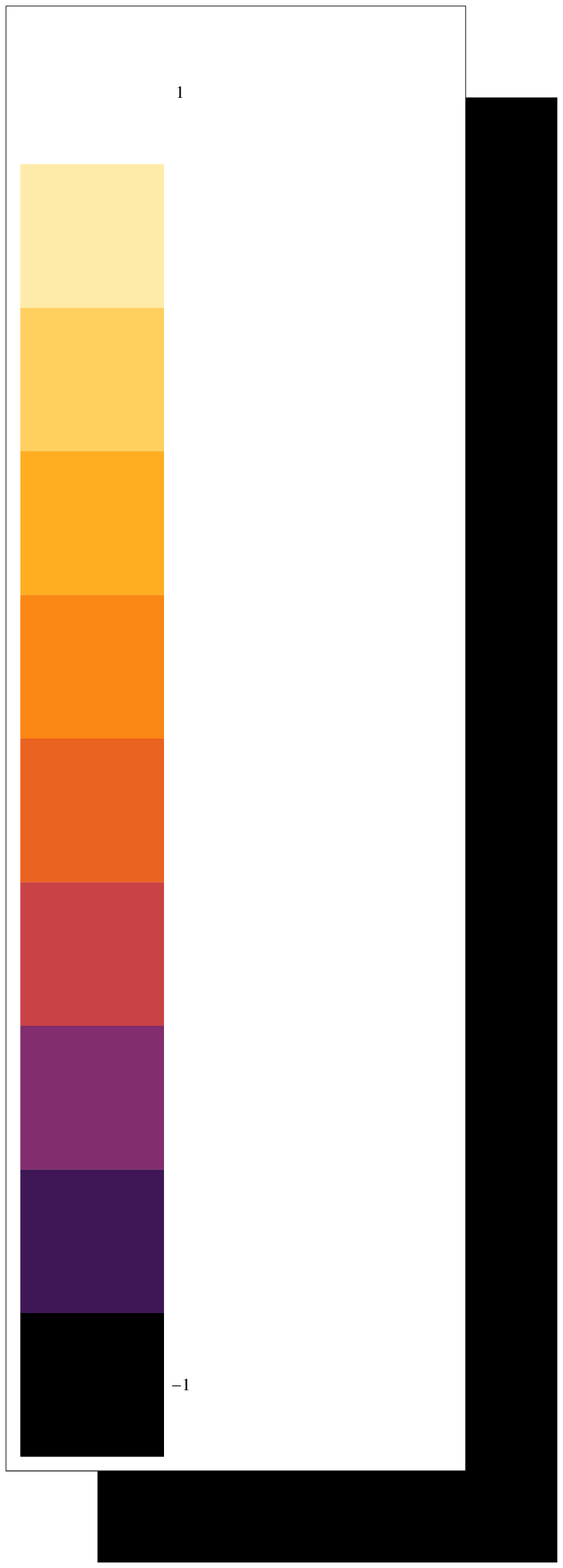}}
\caption{(Color online.) Evolution of the dissipative soliton under the influence of SS effect. (a) negative SS effect ($\sigma_1$=-0.8) and (b) positive SS effect ($\sigma_1$=0.8). Spectral evolutions corresponding to (a) and (b) is shown in (c) and (d), respectively.}
  \label{fig3}
\end{figure*}
 \end{center}
\begin{center}
 \begin{figure*}[htb]
    \subfigure[Input soliton]{\label{4a}\includegraphics[height=3.5 cm, width=4 cm]{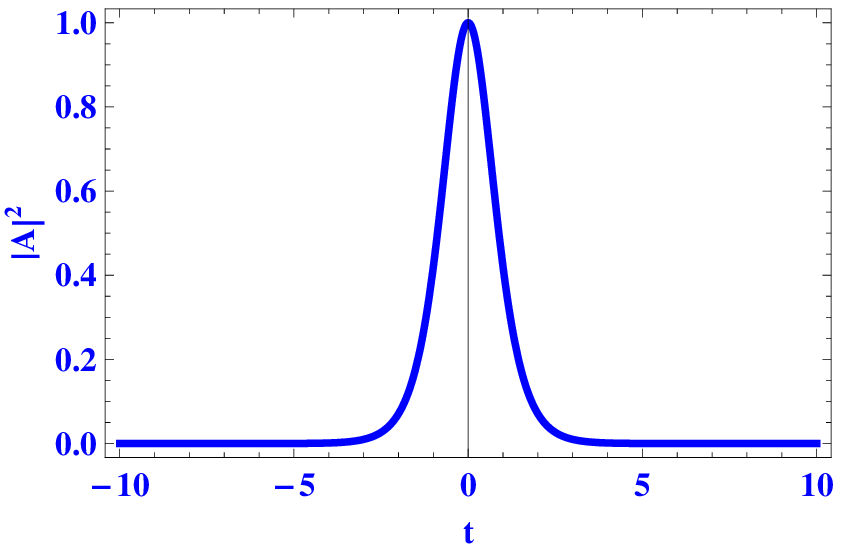}}
    \subfigure[$\sigma_1$=-0.8]{\label{4b}\includegraphics[height=3.5 cm, width=4 cm]{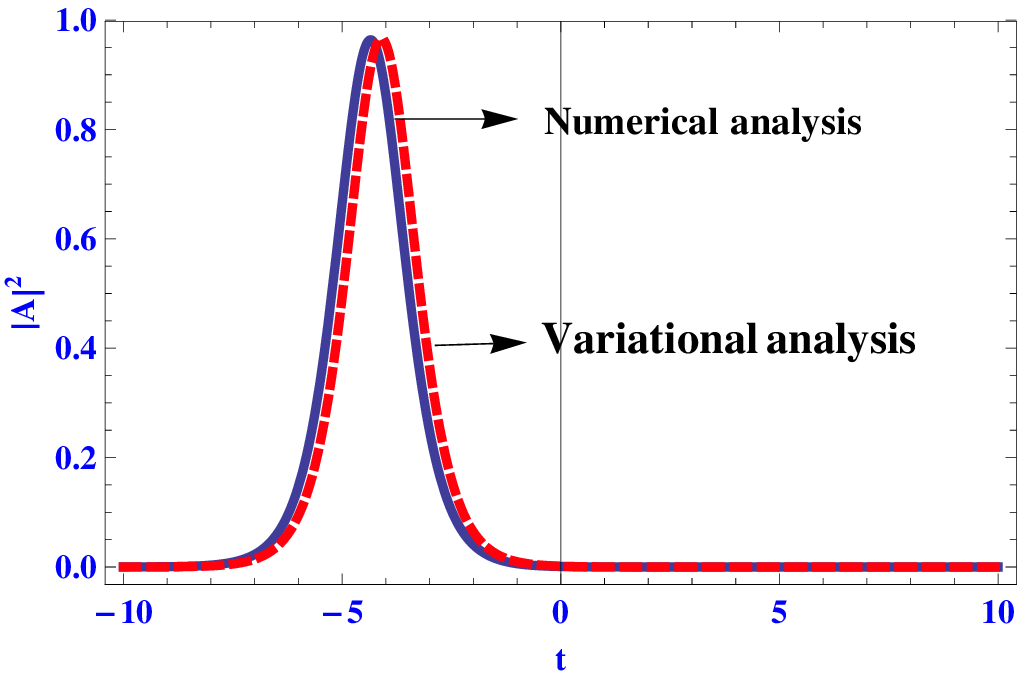}}
    \subfigure[$\sigma_1$=0.8]{\label{4c}\includegraphics[height=3.5 cm, width=4 cm]{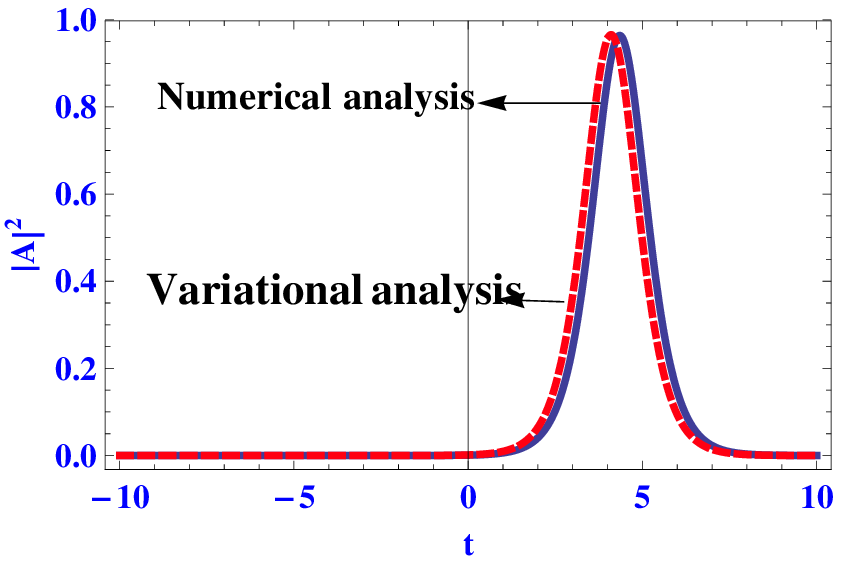}}
      \caption{(Color online.) Temporal shifts of the soliton under the influence of negative and positive SS effects when z=250.}
  \label{d4}
\end{figure*}
 \end{center}
\par
In order to confirm the results of the variational analysis obtained here,  we now solve Eq. (\ref{model}) numerically and present the results in the following. Fig. \ref{fig3} depicts the numerical outcome, which shows the propagation of the DS up to z=250. Figs. \ref{3a} and \ref{3b} correspond to negative ($\sigma_1=-0.8$) and positive ($\sigma_1=0.8$) SS effects. Fig. \ref{d4} shows the input and output intensity profiles of the DS under the influence of SS effects, respectively. It is clear from the figures that the SS effect induces temporal shift during the evolution of the soliton. This SS effect induced temporal shift is toward the leading edge (Figs. \ref{3a} and \ref{4b}) when the SS effect is negative in contrast to the positive SS effect, where the shift occurs toward the trailing edge of the pulse (Figs. \ref{3b} and \ref{4c}). Figs. \ref{3c} and \ref{3d} are spectral evolutions of the soliton corresponding to Figs. \ref{3a} and \ref{3b}, respectively. In Fig. \ref{3c} the spectrum shifts towards the high energy side whereas the spectral shift is toward the red side in Fig. \ref{3d}.
 Also in Figs. \ref{4a} and \ref{4b} we compare the soliton profiles from numerical simulation and variational prediction, respectively at z=250. We can observe that the variational predictions agree
well with numerical simulations. The shape of the pulse remains unchanged under the influence of the SS effect.
\par
In the following section we discuss the impacts of other perturbations considered in Eq. (\ref{per1}) on the self steepening induced temporal shifts and we demonstrate stable dynamics of the DS as a result of the interplay between the higher-order effects.
\subsection{Influence of linear dispersion}
In this section we study the impacts of the higher order linear dispersion such as third and fourth order dispersions on SS induced temporal shift. Initially we discuss the results obtained from the variational analysis.
\begin{figure*}[htb]
    \subfigure[$\sigma_1=0.5$]{\label{sigm}\includegraphics[height=5 cm, width=7 cm]{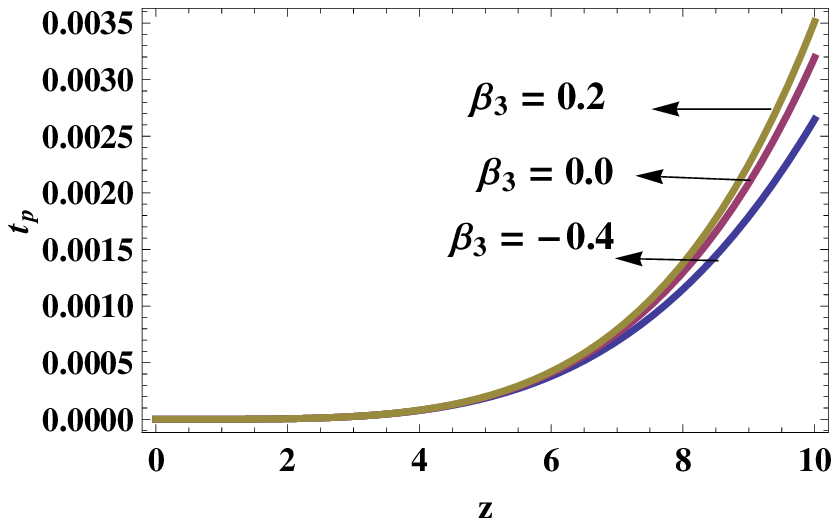}}
    \subfigure[$\sigma_1=-0.5$]{\label{sig}\includegraphics[height=5 cm, width=7 cm]{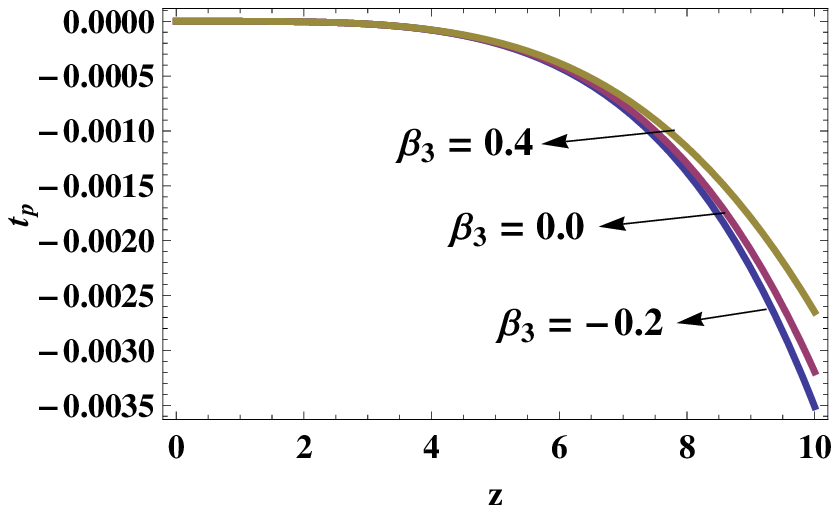}}
      \caption{(Color online.) Impact of TOD on temporal shift $t_p$.}
        \label{bet3}
\end{figure*}
\begin{figure*}[htb]
    \subfigure[$\sigma_1=-0.5$]{\label{bet1}\includegraphics[height=5 cm, width=7 cm]{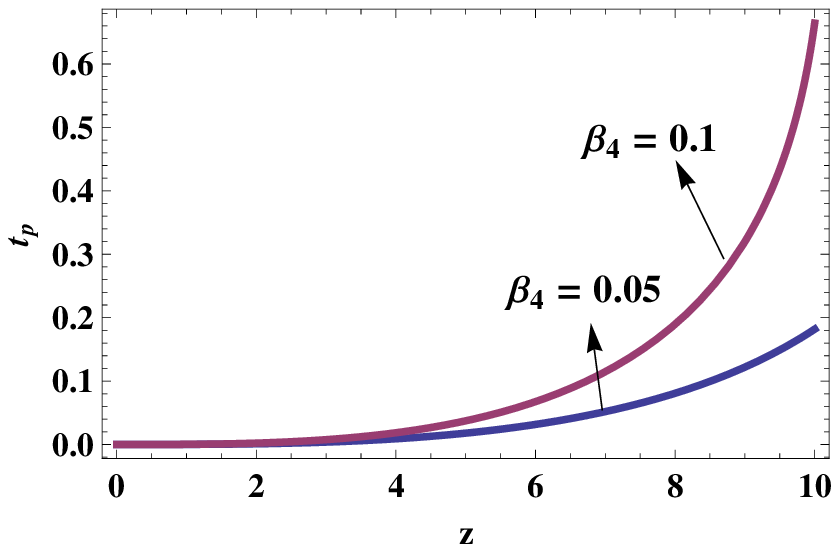}}
    \subfigure[$\sigma_1=0.5$]{\label{bet2}\includegraphics[height=5 cm, width=7 cm]{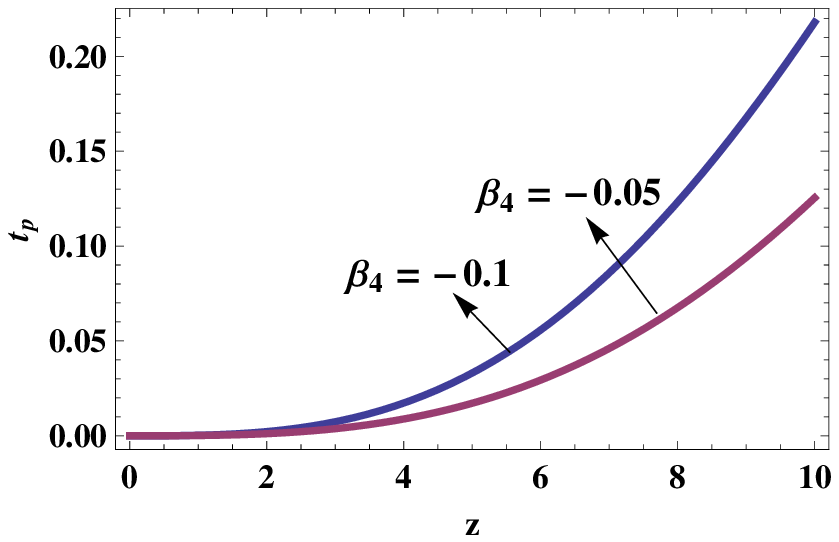}}
      \caption{(Color online.) Impact of the FOD on the temporal shift $t_p$.}
  \label{thr}
\end{figure*}
Fig. \ref{bet3} depicts the impact of TOD on the temporal shift $t_p$. Figs. \ref{sigm} and \ref{sig} correspond to positive and negative SS effects, respectively. It is clear from Fig. \ref{sigm} that when both $\sigma_1$ and $\beta_3$ are positive the peak shift of the soliton towards the trailing edge enhances. On the other hand positive $\sigma_1$ and negative $\beta_3$ suppress the temporal shift. Also in the negative regime of $\sigma_1$ the enhancement and suppression of the peak shift towards the leading edge can be observed when $\beta_3$ is negative and positive, respectively, as shown in Fig. \ref{sig}. However FOD always enhances the SS effect induced temporal shift toward the trailing or leading edges of the pulse, which is clear from Fig. \ref{thr}.
\begin{figure*}[htb]
    \subfigure[$\beta_3=0.2$]{\label{bet3num1}\includegraphics[height=3.5 cm, width=4 cm]{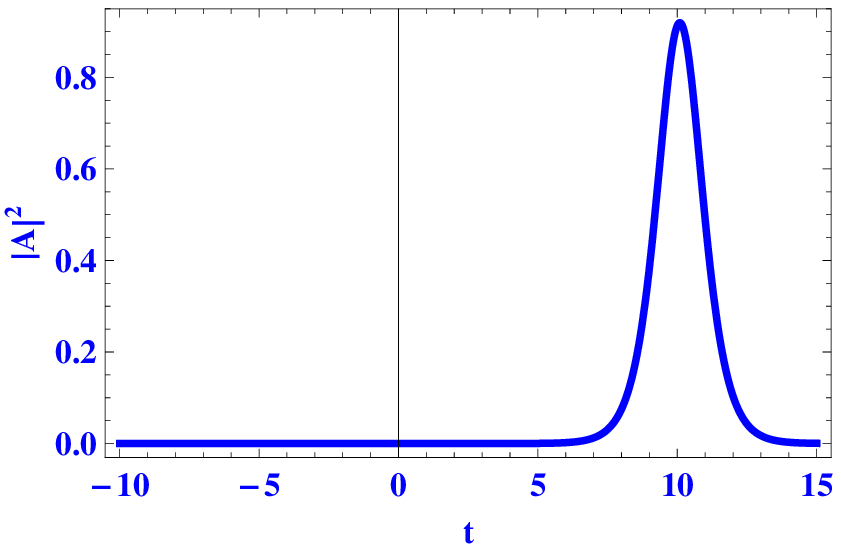}}
    \subfigure[$\beta_3=0.0$]{\label{bet3num2}\includegraphics[height=3.5 cm, width=4 cm]{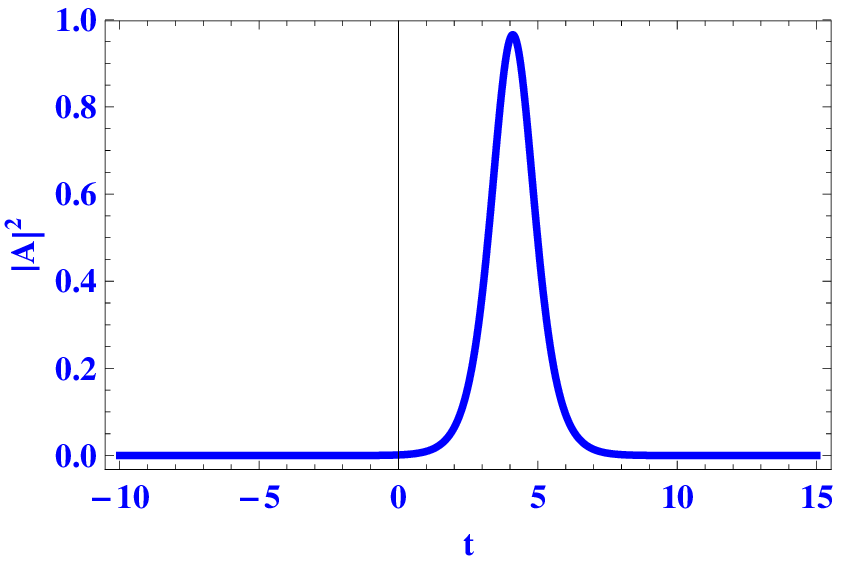}}
    \subfigure[$\beta_3=-0.4$]{\label{bet3num3}\includegraphics[height=3.5 cm, width=4 cm]{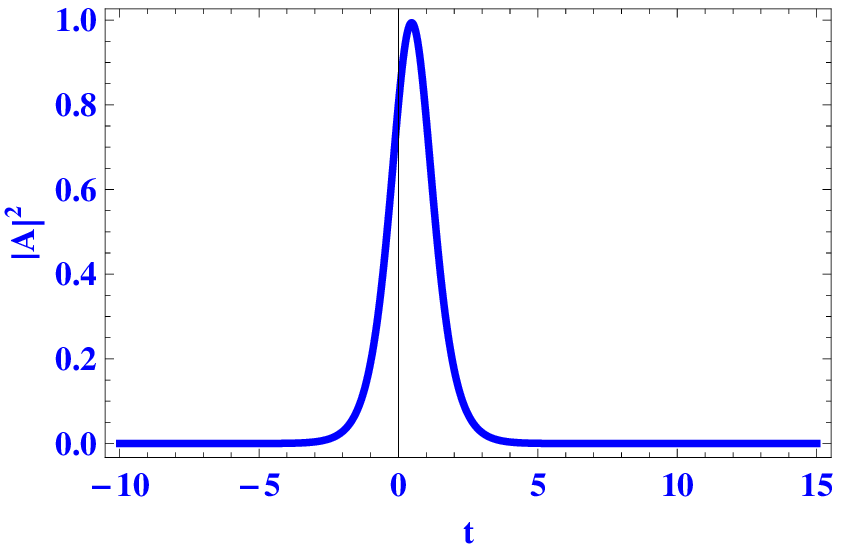}}
      \caption{(Color online.) Intensity profiles of the DS in the positive regime of SS effect ($\sigma_1$=0.8) at z=250 under the influence of different values of $\beta_3$.}
  \label{bet3num}
\end{figure*}
\begin{figure*}[htb]
    \subfigure[$\beta_4=0.0$]{\label{bet4num1}\includegraphics[height=3.5 cm, width=4 cm]{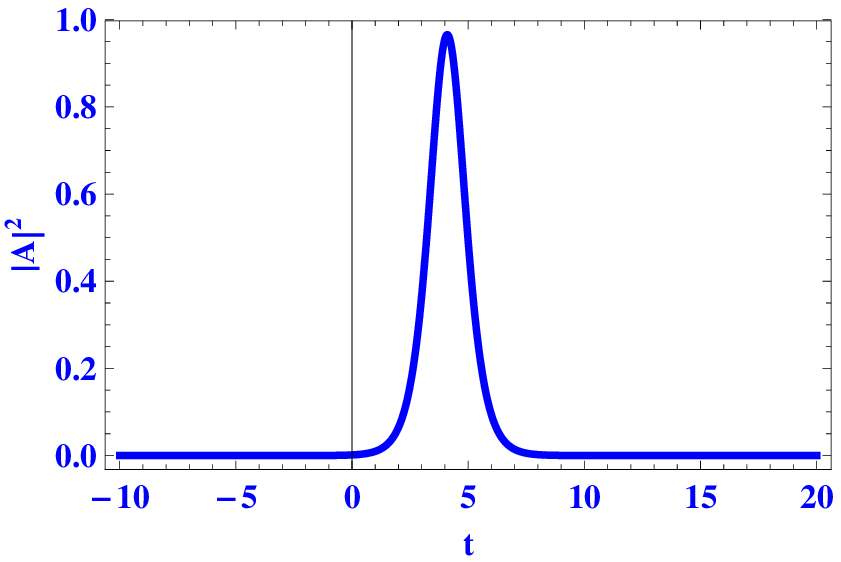}}
    \subfigure[$\beta_4=-0.05$]{\label{bet4num2}\includegraphics[height=3.5 cm, width=4 cm]{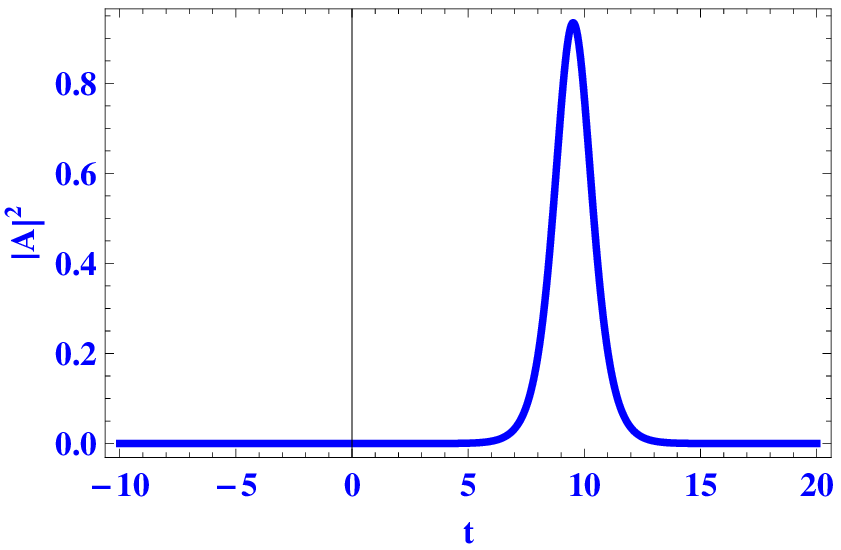}}
    \subfigure[$\beta_4=-0.1$]{\label{Pbet4num3}\includegraphics[height=3.5 cm, width=4 cm]{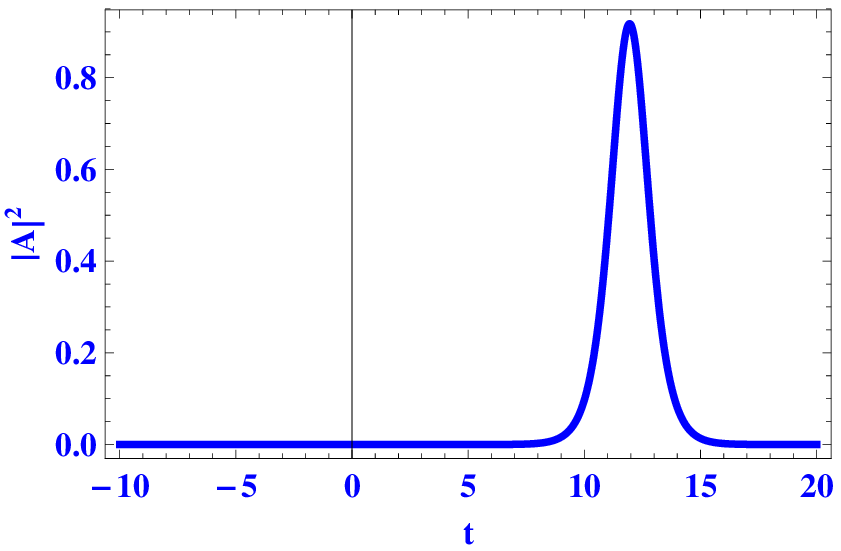}}
      \caption{(Color online.) Intensity profiles of the DS in the positive regime of SS effect ($\sigma_1$=0.8) at z=250 under the influence of different values of $\beta_4$.}
  \label{bet4num}
\end{figure*}
In order to confirm the variational results presented in Figs. \ref{bet3} and \ref{thr} we perform numerical analysis of Eq. (\ref{model}). Fig. \ref{bet3num} depicts the intensity profiles of DS  at z=250 under the influence of different values of $\beta_3$ when $\sigma_1$ =0.8 (a positive value). Fig. \ref{bet3num2} corresponds to the case where $\beta_3$=0, which shows the SS effect induced temporal shift toward the trailing edge of the pulse. Action of positive $\beta_3$ increases the temporal shift along with a reduction of intensity of the profile (Fig. \ref{bet3num1}) \cite{ag}. On the other hand, negative $\beta_3$ tries to bring the soliton peak towards zero temporal shift (Fig. \ref{bet3num3}). As the value of FOD increases the intensity of the profile decreases and it always enhances the SS effect induced temporal shift, which are clear from Fig. \ref{bet4num}.
\subsection{Influence of higher-order nonlinearities}
\begin{figure*}[htb]
    \subfigure[$\sigma_1=-0.5$]{\label{quintic1}\includegraphics[height=5 cm, width=7 cm]{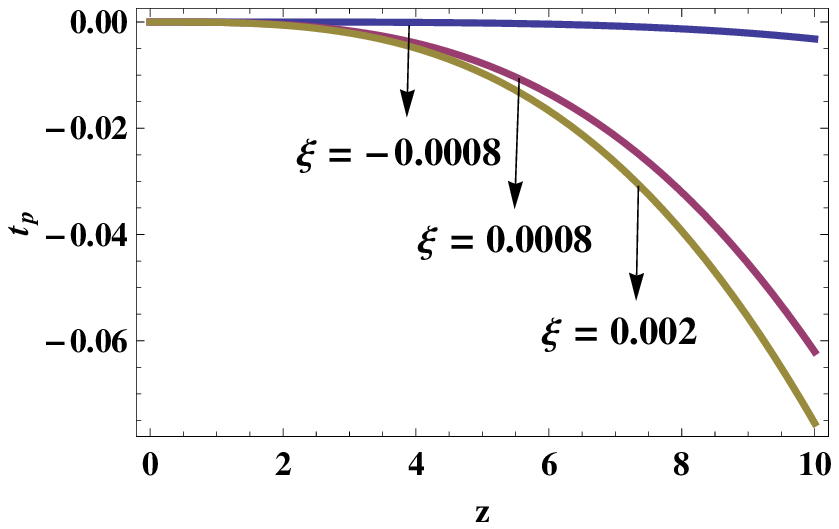}}
    \subfigure[$\sigma_1=0.5$]{\label{quintic2}\includegraphics[height=5 cm, width=7 cm]{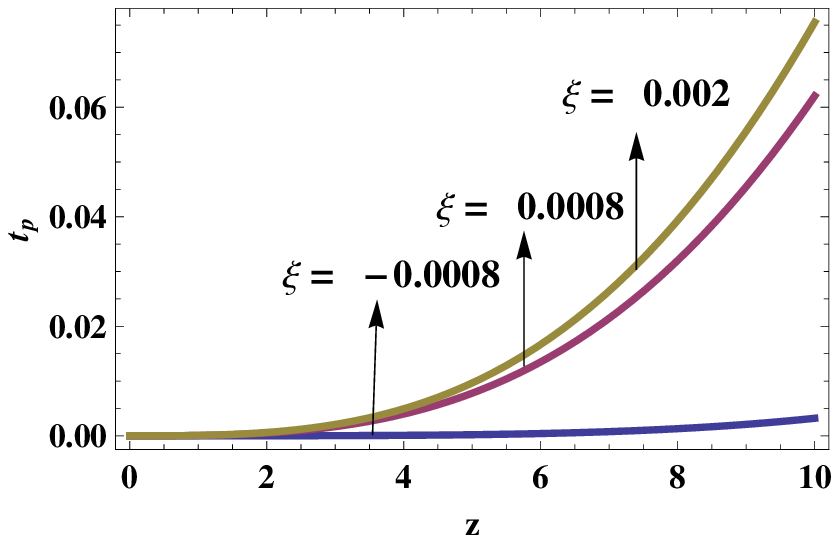}}
      \caption{(Color online.)Variational analysis results which show changes in the SS effect induced temporal shift due to the presence of quintic nonlinearity.}
  \label{quintic1q}
\end{figure*}
Next we discuss the impact of higher-order nonlinear perturbations such as quintic nonlinearity and second order nonlinear dispersion.
\begin{figure*}[htb]
    \subfigure[$\sigma_1=-0.8$]{\label{quintic1}\includegraphics[height=5 cm, width=7 cm]{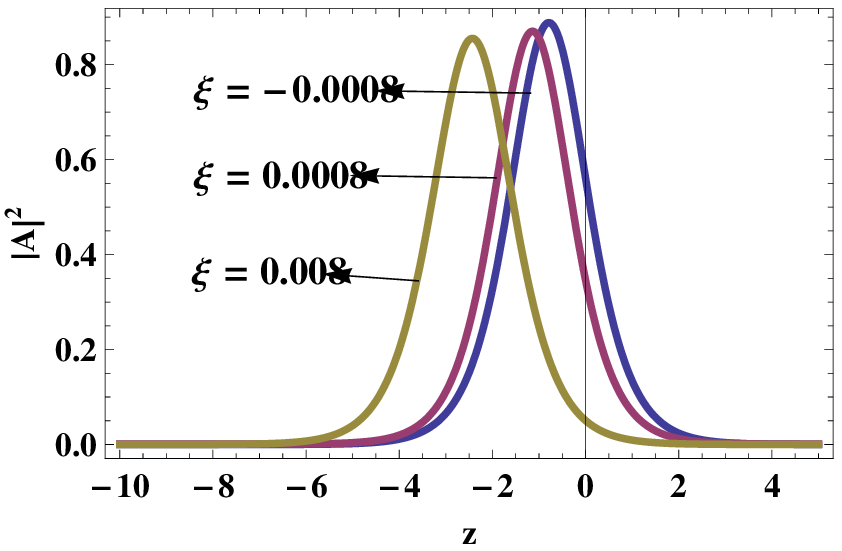}}
    \subfigure[$\sigma_1=0.8$]{\label{quintic2}\includegraphics[height=5 cm, width=7 cm]{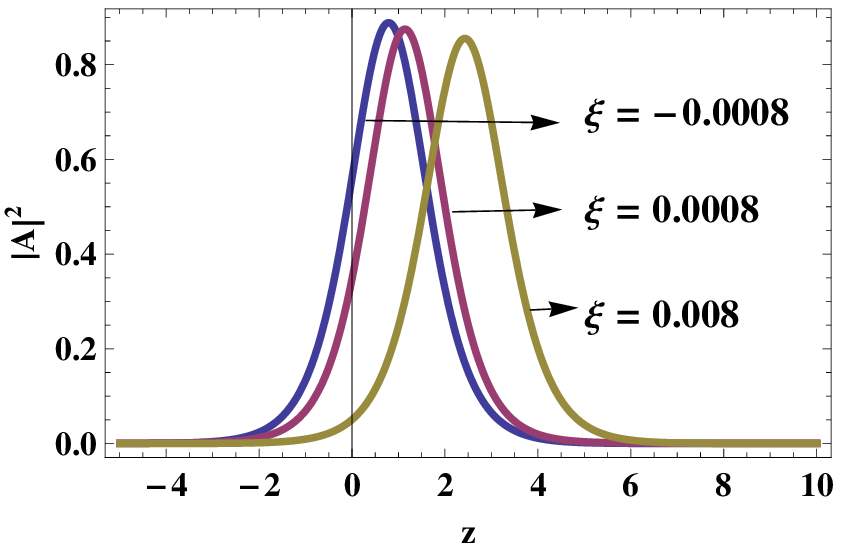}}
      \caption{(Color online.) Impact of QN on SS effect induced temporal shift of soliton.}
  \label{quintic}
\end{figure*}
Figs. \ref{quintic1q} and \ref{quintic} show the variational and numerical results, respectively, which reveal the impact of QN on the SS induced temporal shift. Positive QN coefficient always increases the SS effect induced temporal shift whereas the negative QN coefficient decreases it. Influence of QN is independent of the sign of the SS effect. On the other hand impact of the SOND depends on the sign of the SS coefficient. When the SS effect is positive SOND enhances the temporal shift towards the trailing edge contrary to the case of negative SS effect, where it enhances the shift towards the leading edge of the pulse as depicted in Figs. \ref{snd} and \ref{thre}. Both QN and SOND considerably influence the intensity of the propagating dissipative soliton. Negative QN increases the intensity of the profile whereas positive QN decreases it. On the other hand SOND reduces the intensity of the profile as it is clear from Fig. \ref{thre}.
\begin{figure*}[htb]
    \subfigure[$\sigma_1=-0.5$]{\label{quintic1}\includegraphics[height=5 cm, width=7 cm]{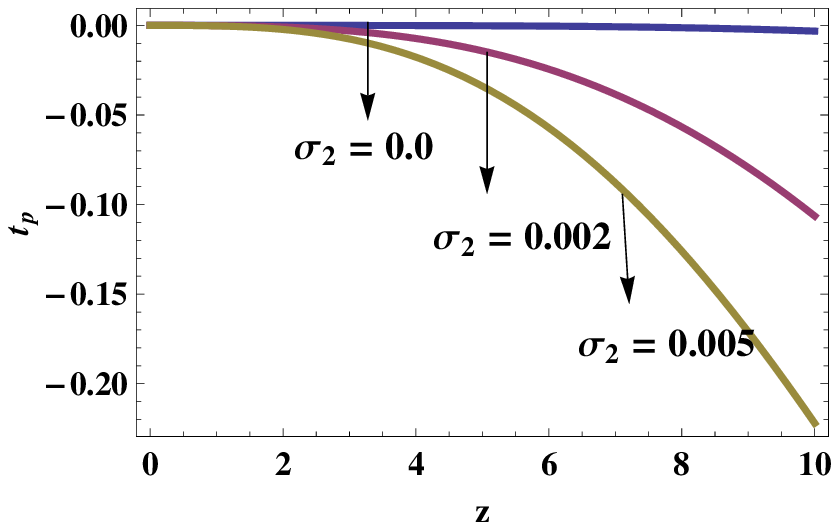}}
    \subfigure[$\sigma_1=0.5$]{\label{quintic2}\includegraphics[height=5 cm, width=7 cm]{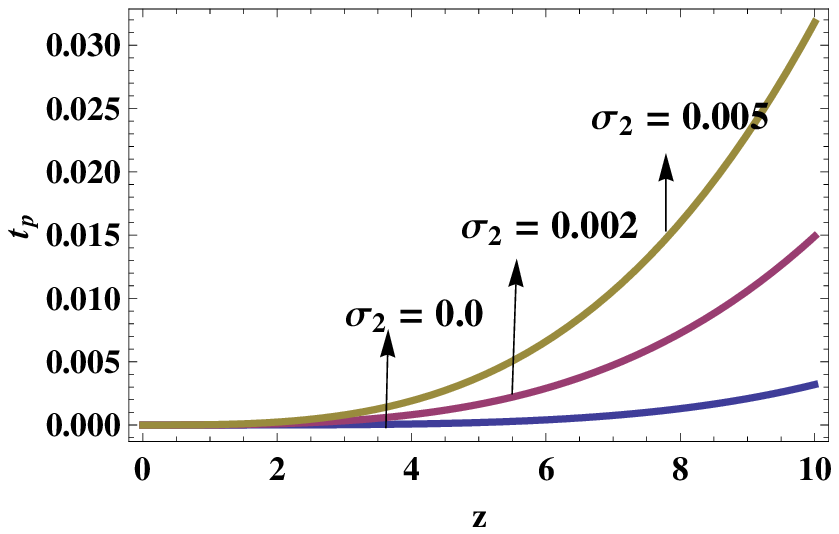}}
      \caption{(Color online.) Variational analysis results which show the influence of SOND on SS effect induced temporal shift.}
  \label{snd}
\end{figure*}
\begin{figure*}[htb]
    \subfigure[$\sigma_2=0.0$]{\label{Nonlinear_normal}\includegraphics[height=3.5 cm, width=4 cm]{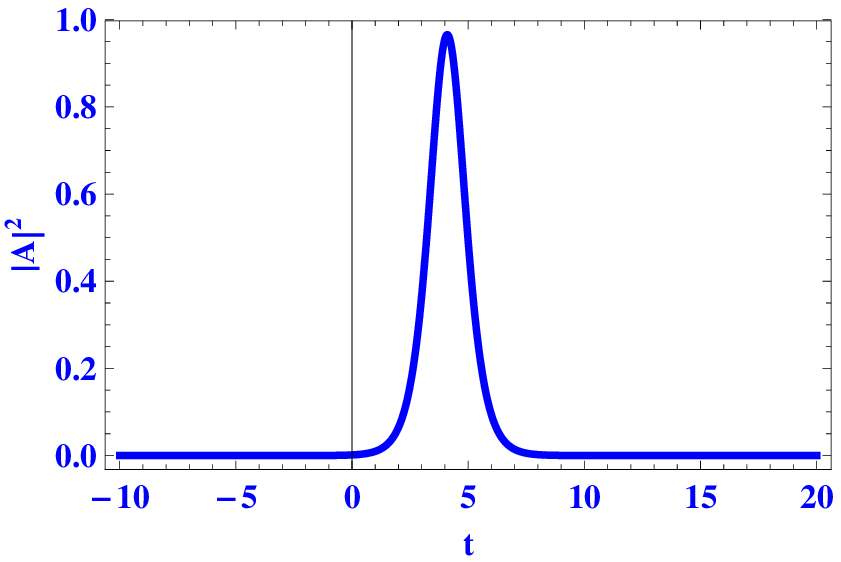}}
    \subfigure[$\sigma_2=0.005$]{\label{PIM_normal}\includegraphics[height=3.5 cm, width=4 cm]{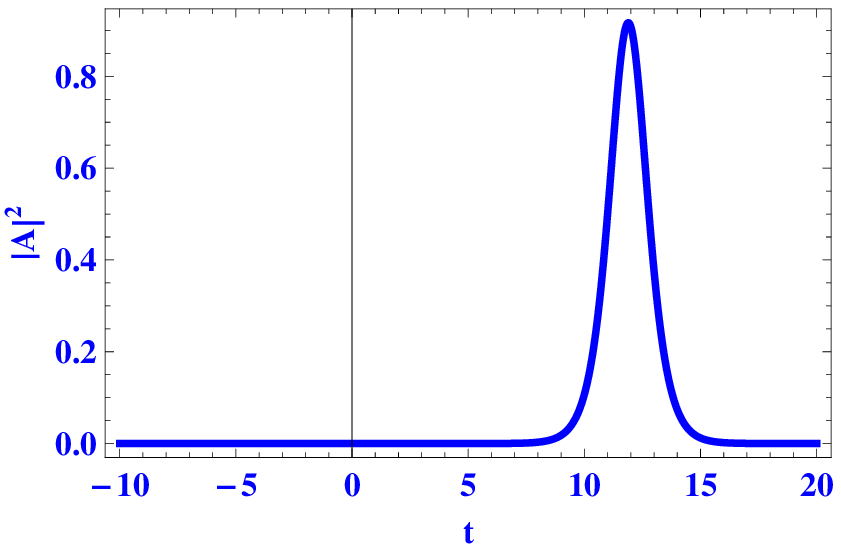}}
    \subfigure[$\sigma_2=0.01$]{\label{PIM_normal}\includegraphics[height=3.5 cm, width=4 cm]{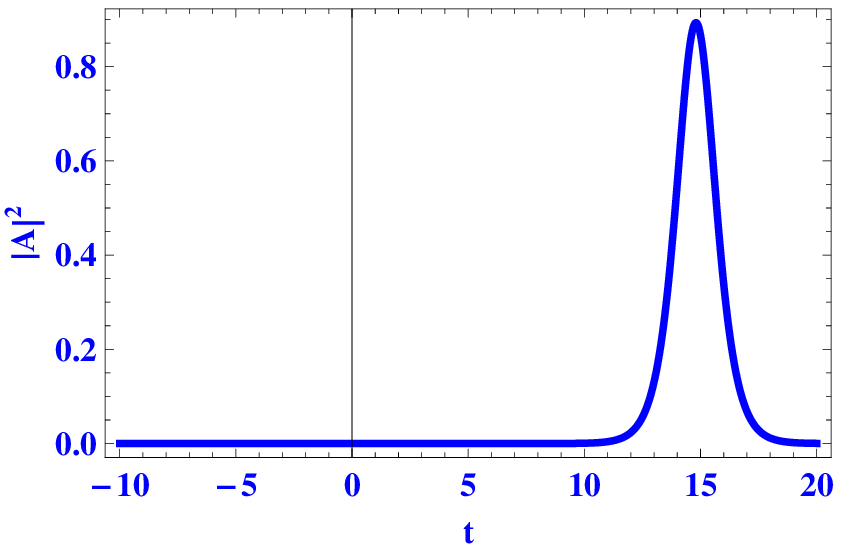}}
      \caption{(Color online.) Impact of the SOND on the SS effect induced temporal shift of soliton when $\sigma_1$=0.8.}
  \label{thre}
\end{figure*}
\subsection{Perturbations balanced soliton propagation}
\begin{figure*}[htb]
    \subfigure[Unperturbed soliton]{\label{sol1}\includegraphics[height=6 cm, width=6 cm]{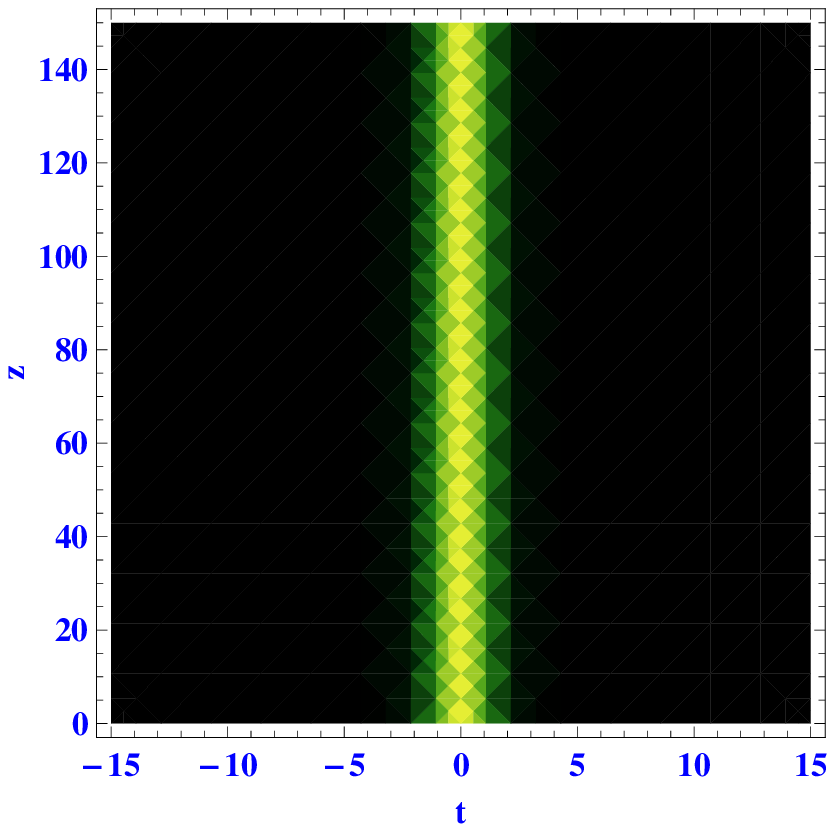}}
    \subfigure[Perturbations balanced soliton]{\label{sol2}\includegraphics[height=6 cm, width=6 cm]{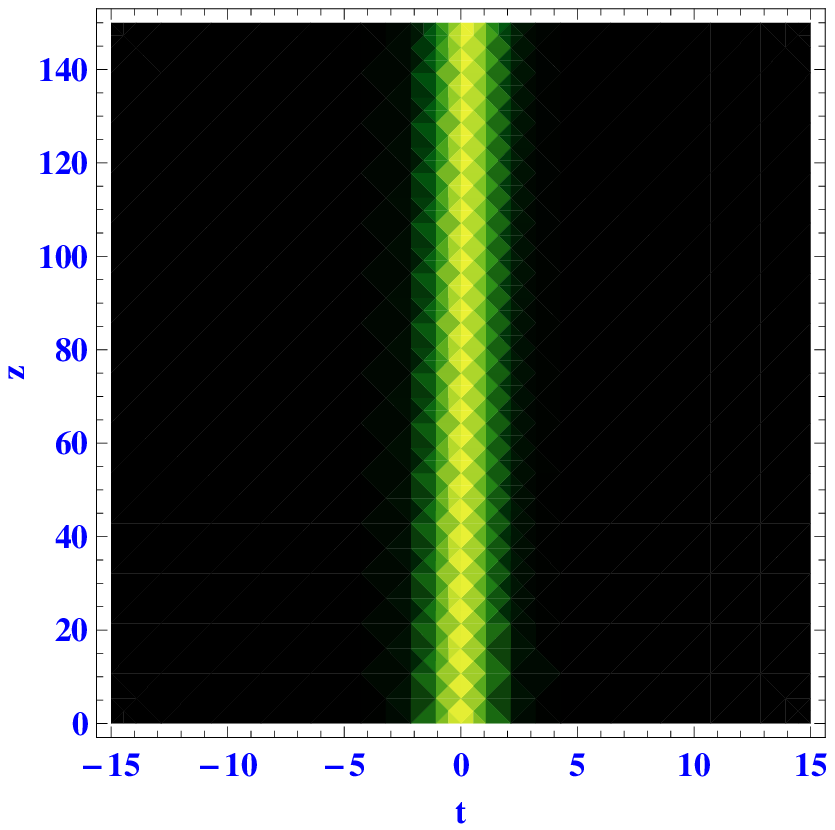}}
      \caption{(Color online.) Unperturbed and perturbation balanced DS propagation in the metamaterials.}
  \label{solipro}
\end{figure*}
It is quite clear from the study on the impact of different perturbations shown in Eq. (\ref{per1}), they considerably influence the dynamics of DS. Now we examine the possibility of stable propagation of the DS in the metamaterials as a result of balancing of the perturbations. Proper combination of higher order effects such as the SS effect and TOD can remove instabilities, which provides stable propagation of the soliton \cite{sofia, Liz, Lu, malo2}. Counteracting perturbations can support stable dynamics of the soliton. The stability analysis of a propagating wave in the metamaterials based on the propagation model given in Eq. (\ref{model}) has been carried out in reference \cite{anse}, which ensures the stability of such waves for appropriate choice of parameters. Fig. \ref{sol1} depicts the DS propagation without the action of any perturbation. The DS propagates without any temporal shift or change in shape. Inclusion of any perturbation defined in Eq. (\ref{per1}) may induce shape change, temporal shift and hence the frequency shift for propagating solitons. Properly choosing  different perturbations, such that they can counteract to form a stable DS. Propagation of such perturbation balanced DS is depicted in Fig. \ref{sol2}. The coefficients of different perturbations used in Fig. \ref{sol2} are $\sigma_1=0.8$, $\sigma_2=0.005$, $\xi=-0.0008$, $\beta_3=-2.0$ and $\beta_4=-0.05$. The perturbations balance each other for this choice of parameters and in the metamaterial stable dynamics of the DS is observed.
\section{Conclusion}
In this theoretical investigation, we have studied the influence of higher-order effects such as TOD, FOD, QN, SS effect and SOND on the dynamics of DS in metamaterials. It is found that the SS effect induces temporal shift and this shift is toward the leading edge when the SS effect is negative in contrast to the positive SS effect, where the shift occurs toward the trailing edge of the pulse. TOD suppresses SS effect induced temporal shift when the sign of TOD coefficient is opposite to the sign of the SS coefficient. On the other hand the FOD always enhances the SS effect induced temporal shift. Positive QN coefficient always increases the SS effect induced temporal shift whereas the negative QN coefficient decreases it. When the SS effect is positive the SOND enhances the temporal shift towards the trailing edge contrary to the case of negative SS effect, where it enhances the shift towards the leading edge of the pulse. We have also found that stable propagation of the DS in the matamaterials can be achieved as a result of balancing of different perturbations for suitable choice of parameters. This study is helpful to form tunable dissipative solitons with potential application in all optical communication system. We believe that our theoretical results will help to stimulate new experiments with metamaterials.
\section{Acknowledgement}
 The work of A.K.S. is supported by the University Grants Commission (UGC), Government of India, through a D. S. Kothari Post Doctoral Fellowship in Sciences. M.L. is supported by DST-SERB through a Distinguished Fellowship (Grant No. SB/DF/04/2017).
 \bibliographystyle{amsplain}

\end{document}